\documentclass[11pt,aps,prb,preprintnumbers,amsmath,amssymb]{revtex4-1}

\usepackage{graphicx}              
\usepackage{dcolumn}               
\usepackage{bm}                    
\usepackage{afterpage}
\usepackage{color}
\usepackage{dsfont}
\usepackage{framed}
\usepackage{algorithm}
\usepackage[noend]{algpseudocode}
\usepackage[output-decimal-marker={.},exponent-product=\cdot]{siunitx}

\def\beq{\begin{eqnarray}}
\def\eeq{\end{eqnarray}}
\def\beqq{\begin{eqnarray*} \color{blue} }
\def\eeqq{\end{eqnarray*}}

\newcolumntype{d}[1]{D{.}{.}{#1}}

\begin{document}

\title{Heat-bath Configuration Interaction: \\An efficient selected CI
 algorithm inspired by heat-bath sampling}

\author{Adam Holmes$^1$}
\email{aah95@cornell.edu}
\author{Norm Tubman$^2$} 
\author{C. J. Umrigar$^1$}
\email{CyrusUmrigar@cornell.edu}
\affiliation{
  $^1$Laboratory of Atomic and Solid State Physics, Cornell University, Ithaca, New York 14853, USA\\
  $^2$University of California, Berkeley, Berkeley, California 94720, USA
}

\begin{abstract}

We introduce a new selected configuration interaction plus perturbation theory
algorithm that is based on a deterministic analog
of our recent efficient heat-bath sampling algorithm. This Heat-bath Configuration
Interaction (HCI) algorithm makes use of two parameters that control the tradeoff between speed and accuracy,
one which controls the selection of determinants to add to a variational
wavefunction, and one which controls the the selection of determinants used to
compute the perturbative correction to the variational energy.
We show that HCI provides an accurate treatment of both static and dynamic correlation
by computing the potential energy curve of the multireference carbon dimer in the cc-pVDZ basis.
We then demonstrate the speed and accuracy of HCI by recovering the
full configuration interaction energy
of both the carbon dimer in the cc-pVTZ basis
and the strongly-correlated chromium dimer in the Ahlrichs VDZ basis, correlating
all electrons, to an accuracy of better than 1 mHa,
in just a few minutes on a single core.
These systems have full variational spaces of $3 \times 10^{14}$ and $2 \times 10^{22}$
determinants respectively.

\end{abstract}
\maketitle

\section{Introduction}
The search for a general, accurate and efficient algorithm for finding approximate
solutions to the quantum many-body problem is one of the major open problems
in electronic structure theory. Methods that work in the basis of Slater determinants
are particularly convenient, because they incorporate the anti-symmetry of the
wavefunction, and classes of one-particle basis sets have been developed that enable
smooth extrapolation to the complete basis set limit. However, the exponential
scaling of the size of the Hilbert space with system size limits deterministic Full
Configuration Interaction (Full CI) calculations to very small systems ($\sim10^{10}$
determinants on a single core).

Stochastic~\cite{booth2009fermion,booth2013towards,ten2013stochastic} and
semistochastic~\cite{petruzielo2012semistochastic}
algorithms have shown great promise in sampling the
Full CI ground and excited states. Deterministic, variational methods
based on tensor networks, such as the
Density Matrix Renormalization Group
(DMRG)~\cite{white1999ab,chan2011density},
are also routinely applied to
strongly-correlated systems. However, (semi)stochastic algorithms
have to introduce an approximation such as the {\it initiator} approximation~\cite{cleland2010communications}
to overcome the infamous Fermion sign problem, 
while computationally tractable tensor network states, such as matrix product states and
tree tensor network states~\cite{nakatani2013efficient},
are inefficient in describing entanglement in molecules that are not
quasi-one-dimensional or tree-like, respectively.

Another well-known approach is selected configuration interaction plus perturbation theory (SCI+PT)
algorithms~\cite{huron1973iterative,buenker1974individualized,schriber2016adaptive,tubman2016deterministic},
which aim to solve the quantum many-body problem within a selected set of determinants.
The first SCI+PT method, known as Configuration Interaction by Perturbatively
Selecting Iteratively (CIPSI)~\cite{huron1973iterative}, iteratively augments a selected space
of determinants as follows. At each iteration,
the ground state within the selected space is obtained,
and the most important components in the first-order perturbation theory correction to that wavefunction
are added to the selected space. After convergence or computational limits
are reached, second-order perturbation theory is performed on the final variational
wavefunction to estimate the Full CI energy.
%
A recent SCI+PT algorithm called Adaptive Sampling CI (ASCI)~\cite{tubman2016deterministic}
accelerates CIPSI by generating connections from only those determinants that have a sufficiently
large amplitude rather than all the determinants in the current variational wavefunction.
In fact the same method had previously been used by two of us for generating trial wavefunctions and deterministic subspaces in the
Semistochastic Full Configuration Interaction Quantum Monte Carlo (S-FCIQMC) algorithm~\cite{petruzielo2012semistochastic,holmes2016efficient}.
ASCI has contributed to renewed interest in SCI+PT methods by reproducing frozen-core DMRG energies for
the challenging Cr$_2$ molecule to within 0.6 mHa in a few CPU hours, albeit in a very small basis,
and provided the motivation for this work.

\begin{figure}
\includegraphics{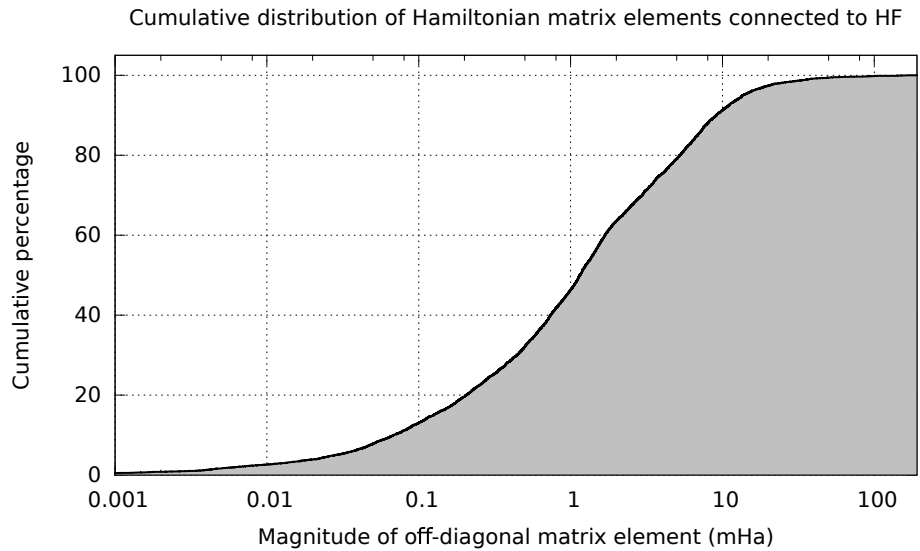}
\protect\caption{
Cumulative distribution of the magnitudes of off-diagonal Hamiltonian matrix elements connected to the Hartree-Fock
determinant for Cr$_2$ at $r=1.5$ \AA, in the Ahlrichs VDZ basis~\cite{schafer1992fully},
where all 48 electrons were correlated.
Hartree-Fock orbitals were used.
All $30\,298$ double-excitation matrix elements larger in magnitude than $10^{-8}$ Ha were included in the cumulative probabilities.
Whereas the largest-magnitude off-diagonal matrix element is 191
mHa, about 97\% of the
matrix elements are at least 10 times smaller
(i.e., smaller than 19.1 mHa), 63\% are at least 100 times smaller,
and some are even more than 1\,000\,000 times smaller
than the largest.
Therefore, generating {\emph all} determinants connected to a reference is an inefficient use of computational resources.
It is more efficient to generate only those determinants that are
connected to the reference by matrix elements larger than a threshold, as described
in the text.
}
\label{off_diag_hist}
\end{figure}

While CIPSI, ASCI, and other SCI+PT algorithms greatly reduce the number of determinants
included in the variational wavefunction, they nevertheless are computationally demanding.
The reason for this is that both SCI+PT steps $-$ identifying
new determinants to add to the selected space and computing a perturbation theory
correction to the energy of the variational wavefunction $-$ require examining \emph{all} of the
determinants connected to a reference determinant by nonzero Hamiltonian matrix elements.
However, as can be seen in Fig.~\ref{off_diag_hist}, the distribution of magnitudes of
off-diagonal matrix elements connected to a reference can span many orders of magnitude.
Determinants that are connected to the reference by very small Hamiltonian
matrix elements will not be added to the space of the variational wavefunction, and
will contribute little to the perturbative correction.
In addition, the coefficients of the reference determinants also vary by orders of magnitude,
and the identification of new determinants to include depends on the product of these coefficients and the matrix elements.
The method we present scans only those determinants for which the above product is above some threshold,
thereby greatly reducing the computer time.

To solve this problem, in this paper, we first introduce a new algorithm for generating determinants
connected to a reference determinant, which we refer to as deterministic heat-bath sampling
because it is a deterministic analog of the (stochastic) efficient heat-bath sampling
algorithm two of us recently developed~\cite{holmes2016efficient}.
Instead of generating \emph{all} single and double
excitations, as is typically done in quantum chemistry algorithms, we instead
generate only those single and double excitations corresponding to Hamiltonian
matrix elements exceeding a threshold $\epsilon$.
The time complexity of our algorithm 
scales only as
the number of determinants that meet the cutoff; no time is wasted on double excitations that do not.

We then incorporate this deterministic heat-bath algorithm into
a new SCI+PT algorithm, which we call Heat-bath CI (HCI), in which both the selection of
new determinants for the variational wavefunction and the computation of the perturbative
correction are greatly accelerated by skipping over determinants connected by small matrix
elements.
HCI is capable of quickly and accurately describing electron correlation.

This paper is organized as follows. In section~\ref{dtm_hb}, we describe our deterministic
heat-bath algorithm for generating only the important determinants
connected to a reference determinant. In section~\ref{sel_ci}, we describe HCI, our new SCI+PT
algorithm, which is based on our deterministic heat-bath algorithm.
In section~\ref{results}, we apply HCI to the potential energy curve of the carbon dimer and to the
chromium dimer, to benchmark its accuracy against highly accurate 
algorithms such as Full CI and DMRG.
Finally, in section~\ref{ongoing_research} we describe our current research directions.


\section{Deterministic Heat-Bath ``Sampling''}
\label{dtm_hb}

Here we describe our algorithm for generating only those determinants
that are connected to a single reference determinant by Hamiltonian matrix elements whose
magnitudes exceed a cutoff $\epsilon$. We call this a deterministic heat-bath
algorithm because it is the deterministic analog of the efficient heat-bath
sampling algorithm recently developed by two of us~\cite{holmes2016efficient}.
Stochastic efficient heat-bath sampling enables one to efficiently sample the
determinants connected to a reference determinant according to an approximate heat-bath distribution,
i.e., with probability approximately proportional to the absolute value of the Hamiltonian matrix element connecting
the target determinant to the reference determinant.

While configuration state functions (CSFs) 
and non-orthogonal Slater determinants certainly have their
advantages~\cite{braida2011quantum,fracchia2012size,tsuchimochi2016black},
we instead choose to work in the space
of orthogonal determinants because the quantum chemical Hamiltonian takes on a
very simple structure~\cite{knowles1984new}, which we use to our advantage.
The inspiration for this deterministic heat-bath algorithm comes from
two observations on the form of the quantum chemical Hamiltonian:
\begin{enumerate}
\item Most of the nonzero off-diagonal elements of the Hamiltonian are double excitations.
\item The magnitudes of double excitations (not the signs) depend only on the four orbitals
that change occupancy during the excitation and not on any other orbitals.
\end{enumerate}
Because of observation 1, we will get the most gain by accelerating
the algorithm for generating double excitations. Because of observation
2, we can organize and store the magnitudes of double excitations any way we choose
once and for all at the beginning of a run.

Let $\left|H\left(rs\leftarrow pq\right)\right|$ denote the magnitude
of a double excitation in which electrons in spin-orbitals $p$ and
$q$ excite to spin-orbitals $r$ and $s$. As noted in observation
2, the magnitude of this matrix element depends only on the spin-orbitals
$\left\{ p,q,r,s\right\}$,
independent of which other spin-orbitals are occupied.

The deterministic heat-bath algorithm has two parts: a setup routine, to be called once
at the beginning of the run, and a routine for generating determinants connected to a
reference determinant by matrix elements with magnitude larger than a threshold,
that can be called any time the important determinants connected to a reference are needed.
This latter routine will be a crucial component of our Heat-bath CI algorithm,
described in section~\ref{sel_ci}.

\subsection{Setup}

Store the set of double excitations as follows: For each pair of orbitals
$\left\{ p,q\right\} $, store a list of triplets $\left\{ r,s,\left|H\left(rs\leftarrow pq\right)\right|\right\} $,
one triplet for each distinct pair of orbitals $\left\{ r,s\right\} $
that do not include $\left\{ p,q\right\} $, sorted by $\left|H\left(rs\leftarrow pq\right)\right|$
in decreasing order.
Also, store 
the maximum magnitude of a double
excitation, $H_{{\rm max}}^{{\rm doub}}$.

This setup has time complexity $\mathcal{O}\left(M^{4}\log M\right)$
and space complexity $\mathcal{O}\left(M^{4}\right)$, where $M$
is the number of orbitals.

\subsection{Generating connected determinants}

When generating determinants connected to a reference determinant, the usual strategy
employed by other quantum chemistry methods such as CISD, MP2,
CIPSI and ASCI is to generate \emph{all} single and double excitations from the reference determinant.
However, we do not do this because
many of the matrix elements connecting these excited determinants
to the reference determinant are very small, and the time
required to generate those excitations may be better spent elsewhere.

Instead, we introduce a threshold, $\epsilon$, and generate only
those determinants that are connected to the reference determinant by Hamiltonian
matrix elements that are larger in magnitude than $\epsilon$, as
follows:
\begin{enumerate}
\item \textbf{Generate only those double excitations that exceed $\epsilon$.}
If $\epsilon>H_{{\rm max}}^{{\rm doub}}$, no double excitations are generated.
Otherwise, loop over all pairs of occupied orbitals $\left\{ p,q\right\} $.
For each pair, look up the stored list of triplets $\left\{ r,s,\left|H\left(rs\leftarrow pq\right)\right|\right\} $,
omitting those in which $r$ or $s$ is occupied,
until a triplet is reached for which
$\left|H\left(rs\leftarrow pq\right)\right|<\epsilon$.
\item \textbf{Generate all single excitations, then discard those that are
smaller than $\epsilon$.} 
Loop over all occupied orbitals $p$. For each $p$, loop over all orbitals $\left\{r\right\}$ in the same
irreducible representation, omitting those in which $r$ is occupied in the reference determinant.
Compute $H\left(r\leftarrow p\right)$,
the matrix element corresponding to the single excitation in which
an electron moves from orbital $p$ to orbital $r$. If $\left|H\left(r\leftarrow p\right)\right|<\epsilon$,
discard the single excitation $\left(r\leftarrow p\right)$.
The cost of generating the single
excitations is $\mathcal{O}\left(N^2M\right)$, where $N$ and $M$ are the numbers of electrons
and orbitals, respectively,
since each single excitation matrix element takes $\mathcal{O}\left(N\right)$ time to compute.
\end{enumerate}
This algorithm has a time complexity of 
$\mathcal{O}\left(N_{{\rm con}}^{\epsilon}+N^2M\right)$, 
where $N_{{\rm con}}^{\epsilon}$ is the number of determinants
connected to the reference by matrix elements that are larger in
magnitude than $\epsilon$.
No time is wasted on those doubly-excited determinants connected to the reference
by matrix elements smaller in magnitude than $\epsilon$.

By varying the threshold $\epsilon$, one can vary between ``accurate
but slow'' (small $\epsilon$) and ``less accurate but fast''
(large $\epsilon$), depending on the demands of the system being
studied. When $\epsilon=0$, this algorithm reduces to the standard
algorithm of generating \emph{all} determinants connected to a reference,
i.e., ``accurate but probably unnecessarily slow.''

\section{Heat-bath Configuration Interaction}
\label{sel_ci}

In this section, we apply the deterministic heat-bath sampling algorithm of section~\ref{dtm_hb}
to invent an SCI+PT algorithm, which we call Heat-bath CI (HCI).
There are two stages: generating the variational wavefunction and energy,
and computing the perturbative energy correction. We formulate our algorithm
such that it has only two parameters, one for each stage.

\subsection{Generating the variational wavefunction}

Like CIPSI and ASCI, HCI generates a variational wavefunction using an iterative process
in which at each iteration we diagonalize in the selected space, and then
add new determinants to the space. In order to identify the new determinants $\left\{D_k\right\}$ to add,
both CIPSI and HCI choose those determinants that are
most ``important" according to some importance measure $f\left(D_k\right)$,
from among the set of determinants
connected to the current selected space by nonzero Hamiltonian matrix elements.
(ASCI, in common with S-FCIQMC~\cite{petruzielo2012semistochastic,holmes2016efficient},
chooses new determinants from only those connected to a truncated
subspace of the current selected space, but
is otherwise identical to CIPSI.)
However, the importance measures used by CIPSI and HCI are different.
CIPSI uses the first-order
perturbation theory estimate of the coefficients, i.e.,
\beq
\label{1pt}
f_{\rm CIPSI}\left(D_k\right)=\left|c_k^{\left(1\right)}\right|&=&\left|\frac{\sum_i H_{ki} c_i}{E_0-H_{kk}}\right|,
\eeq
whereas HCI uses the simpler measure,
\beq
\label{f_hci}
f_{\rm HCI}\left(D_k\right)&=&\max_i\left(\left|H_{ki} c_i\right|\right).
\eeq
which completely eliminates the need to query unimportant determinants, as discussed shortly.

This measure is justified as follows.
As previously demonstrated in Fig.~\ref{off_diag_hist}, for fixed $i$,
the range of values that $\left|H_{ki}\right|$ can take
spans many orders of magnitude. The range of possible values of
$\left|H_{ki}c_i\right|$ is even larger, since the coefficients $\left\{c_i\right\}$
can also vary widely in magnitude. On the other hand, the denominators in
Eq.~\ref{1pt} do not vary as widely, and are unlikely to range in value by even
one order of magnitude.
This is particularly true in later iterations, since by then, determinants with
low diagonal matrix elements $H_{kk}$ have already been included in the wavefunction.
Therefore, most of the variation in Eq.~\ref{1pt}
is dominated by $\max_i\left(\left|H_{ki} c_i\right|\right)$. 

\begin{figure}
\includegraphics{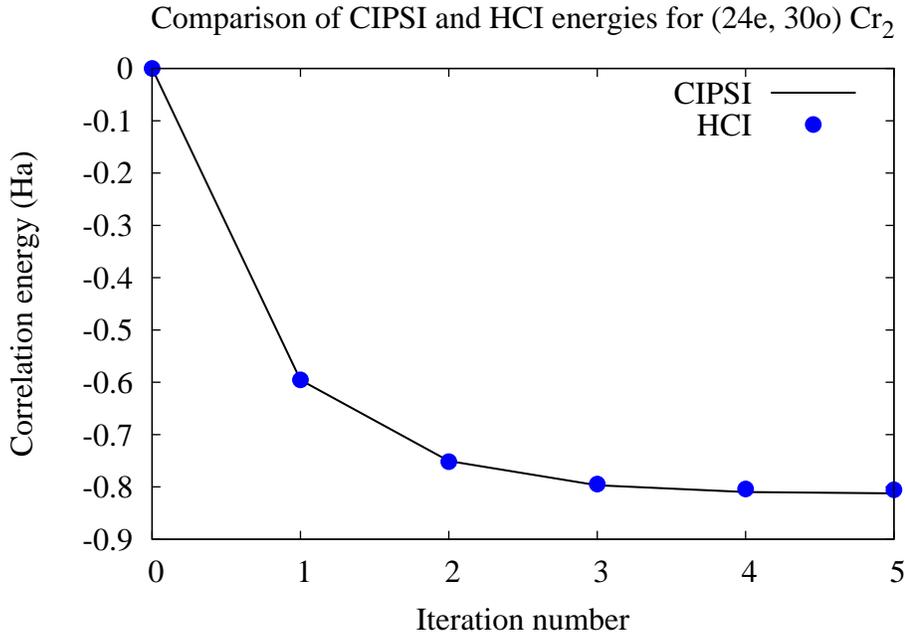}
\protect\caption{
Comparison of the variational correlation energies obtained using the CIPSI and HCI
importance measures
(Eqs.~\ref{1pt} and~\ref{f_hci}), for the chromium dimer at $r=1.5$ \AA, in the
Ahlrichs VDZ basis~\cite{schafer1992fully}. Natural orbitals from a CASSCF(12,12) were used,
and Mg cores were frozen. The HCI energies were
obtained using $\epsilon_1=1$ mHa, and it converged in 5 iterations. The CIPSI
energies were obtained by adding the same number of determinants that HCI
added each iteration, but the new determinants were chosen
by searching all determinants connected to the previous iteration's ground state
and choosing those with the largest importance according to Eq.~\ref{1pt}.
Although the two importance measures, $f_{\rm CIPSI}$ and $f_{\rm HCI}$, are
different, the variational wavefunctions they produce iteratively are very similar in
energy.
The variational energy from HCI is 7 mHa higher at the final iteration.
After the perturbative correction the difference in the energies is much smaller than
this and can be of either sign.
}
\label{cipsi}
\end{figure}

Because the variations in both importance measures are dominated by
variations in $\max_i\left(\left|H_{ki} c_i\right|\right)$, they are likely to
yield very similar rankings of the candidate determinants by importance.
Recall that both CIPSI and HCI use their respective
importance measures only to divide the determinants into two groups (important
ones to add to the selected space, and unimportant ones to discard). Small
differences in ordering make no difference unless they cause determinants to move
between the two groups, and those that switch groups are likely to carry a small
weight after diagonalization anyway. Fig.~\ref{cipsi} shows a comparison of
two 5-iteration runs, in which the same numbers of determinants were added
each iteration, but with the new determinants selected according to the two
different importance measures. As expected, the two importance measures produce
variational wavefunctions that are very similar in energy on each iteration.

This new measure for selecting determinants (Eq.~\ref{f_hci}) enables the
computational cost of HCI to be much less than that of CIPSI for two reasons.
First, we examine only those determinants that are connected to a determinant in
the reference
wavefunction by Hamiltonian matrix elements larger than a threshold, which avoids
wasting time on determinants that are unlikely to be important components of the
first-order correction.
Second, since all of these determinants are connected strongly enough to the reference
to have already met a threshold, we add all of them to the selected space, and bypass the
costly computation of first-order perturbation theory estimates of the coefficients altogether.

To accomplish this,
we use our deterministic heat-bath sampling algorithm, described in section~\ref{dtm_hb}.
This stage of the algorithm requires one parameter, $\epsilon_1$.
On the first iteration, we start with a selected space consisting of only the
Hartree-Fock determinant. We iterate as follows.
\begin{enumerate}
\item Find the lowest eigenvector of the Hamiltonian in the selected space,
and denote the determinant coefficients as $\left\{c_i\right\}$.
\item Find all determinants $\left\{D_k\right\}$ outside of the selected space for which
$\left|H_{ki}c_i\right|>\epsilon_1$ for at least one determinant $D_i$ in the selected space.
\item Add those determinants to the selected space.
\item Repeat steps 1-3 until 
an iteration is reached in which the number of new determinants is less than 1\% of
the number of determinants already selected. 

\end{enumerate}

Step 2 can be accomplished efficiently as follows: Iterate over all determinants $\left\{D_i\right\}$ in the selected
space. For each, use the deterministic heat-bath sampling algorithm to generate
all determinants connected to $D_i$ by Hamiltonian matrix elements larger than
$\epsilon=\epsilon_1/\left|c_i\right|$.
Then, the lists of connected determinants are merged (duplicates are removed).

\subsection{Perturbative correction}

The second-order perturbative correction to the variational energy
$E^{\left(0\right)}$ is given by
\beq
\Delta E^{\left(2\right)}&=&\sum_k \frac{\left(\sum_i H_{ki} c_i\right)^2}{E^{\left(0\right)}-H_{kk}},
\eeq
where the sum on $k$ runs over all determinants outside of the selected space,
that are connected to at least one determinant in the selected
space by a nonzero Hamiltonian matrix element.

As previously mentioned, the magnitudes of $\left\{H_{ki}\right\}$ can span
many orders of magnitude, so many terms in the sum above
are very small. We therefore introduce an approximate perturbation theory expression that makes use of
one more parameter, $\epsilon_2$:
\beq
\Delta E^{\left(2\right)}
&\approx&\sum_k \frac{\left(\sum_i^{\left(\epsilon_2\right)} H_{ki} c_i \right)^2}{E^{\left(0\right)}-H_{kk}},
\eeq
where $\sum^{\left(\epsilon_2\right)}$ denotes a sum in which all terms in the sum smaller in magnitude than
$\epsilon_2$ are removed. In other words, we approximate the sum on $i$ in the numerator by skipping over the small contributions
$H_{ki} c_i$ for which $\left|H_{ki}c_i\right|<\epsilon_2$.

This stage can be accomplished efficiently using the deterministic heat-bath algorithm of Section~\ref{dtm_hb} as
follows. Iterate over all determinants $\left\{D_i\right\}$ in the variational wavefunction. For each one,
generate all determinants $\left\{D_k\right\}$ connected to $D_i$ by matrix elements larger than
$\epsilon=\epsilon_2/\left|c_i\right|$, and store both their labels and corresponding values of
$H_{ki}c_i$. Once all have been generated, merge the lists, so that the sums $\sum_i H_{ki}c_i$
can be computed before they are squared.

Note that this na\"ive approach requires storing all of the determinants that contribute
to the second-order energy, so it will be too expensive when that number is large.
We are currently developing an alternative method with a much smaller storage requirement,
which will enable us to use much larger variational determinant expansions.
Nevertheless, this simple approach suffices to obtain all the results in this
paper in a few minutes on a single core.

By varying $\epsilon_2$, we can obtain approximate values for the second-order energy correction quickly, without wasting any time on the small terms that are excluded from the sum. We can also extrapolate to the limit $\epsilon_2\rightarrow 0$, even when computing
the exact expression ($\epsilon_2=0$) would be prohibitively expensive.

\subsection{Context within existing quantum chemistry algorithms}

We now place HCI in the context of other quantum chemistry algorithms.
An HCI run is specified by two parameters: $\epsilon_1$, which controls
which determinants will be included in the variational wavefunction,
and $\epsilon_2$, which controls the accuracy of the perturbative correction to the variational
energy. We therefore denote a particular instance of HCI as HCI($\epsilon_1$,$\epsilon_2$).

When $\epsilon_1$ is larger than the magnitudes of all off-diagonal Hamiltonian matrix
elements connected to the Hartree-Fock (HF) reference, no determinants will be added to the variational
wavefunction. In that case, the variational wavefunction will reduce to the HF determinant.
When the variational wavefunction is HF, the perturbative correction will yield the
second-order Epstein-Nesbet perturbation theory (EN-PT) energy correction 
if $\epsilon_2=0$, and will yield zero if $\epsilon_2$ is larger
than the magnitudes of all off-diagonal Hamiltonian matrix elements connected to HF
(since no terms would be included in the sum).

When $\epsilon_1=0$, all determinants will (after many iterations) be added to the selected
space, resulting in a variational wavefunction equal to the Full CI (FCI) ground state. In that
case, the perturbative correction will be zero, no matter what value $\epsilon_2$ has,
since there will be no other determinants left.

In summary,
\beq
{\rm HCI}\left(\epsilon_1,\epsilon_2\right)&=&\begin{cases}
{\rm HF},&{\rm if}\quad \epsilon_1\ge H_{{\rm max}}^{{\rm doub}}\quad{\rm and }\quad\epsilon_2\ge H_{{\rm max}}^{{\rm doub}};\\
\\
{\rm EN\mbox{-}PT},&{\rm if}\quad \epsilon_1\ge H_{{\rm max}}^{{\rm doub}}\quad{\rm and }\quad\epsilon_2=0;\\
\\
{\rm FCI},&{\rm if}\quad\epsilon_1=0.
\end{cases}
\eeq

HCI can thus be seen as a generalization of HF, EN-PT, and FCI, that is more flexible than
any of them in enabling a tradeoff between the speed and accuracy of its ground state
energy calculations. As we shall see in the following section,
FCI-quality results can be obtained at a significantly reduced
cost by choosing $\epsilon_1$ and $\epsilon_2$ appropriately.


\section{Results}
\label{results}
We applied our HCI algorithm to two systems, the carbon dimer and the chromium dimer.
All integrals, coupled cluster calculations, and orbital rotations were computed using the \textsc{Molpro}
quantum chemistry software package~\cite{werner2012molpro}.

\subsection{Carbon dimer}
We first applied HCI to the potential energy surface of the carbon dimer, a system known
to be of highly multireference character, even at equilibrium geometry.
We used the cc-pVDZ basis set~\cite{dunning1989gaussian} (28 spatial orbitals) and correlated all 12 electrons.
The Full CI space consists of about $2\times 10^{10}$ Slater determinants.
We compared our results to recently-published Full CI
values~\cite{sharma2015multireference}, since they
were tabulated for a large part of the binding curve.
By running tests at equilibrium geometry at various values of $\epsilon_1$ and $\epsilon_2$,
we found that using $\epsilon_1=1$ mHa and
$\epsilon_2=30$ $\mu$Ha gave a ground-state energy converged to within
1 mHa of the Full CI energy. We then used these parameter values
for the whole curve.

\begin{figure}
\includegraphics{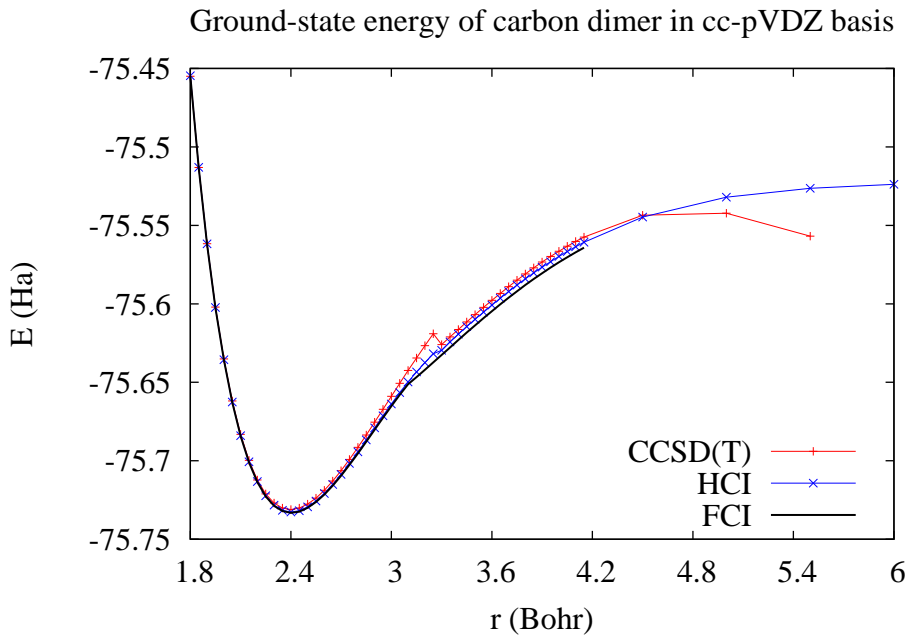}
\protect\caption{
Binding curve of C$_2$ in a cc-pVDZ basis obtained from HCI($\epsilon_1=1$ mHa, $\epsilon_2=30$ $\mu$Ha)
compared to CCSD(T) (computed using \textsc{Molpro}) and Full CI~\cite{sharma2015multireference}.
CCSD(T) is good at describing dynamic correlation but poor at describing static correlation,
so while it gives good energies near equilibrium,
it can't describe bond breaking well. On the other hand, HCI gives
good energies along the whole dissociation curve, indicating that it
accurately describes both static and dynamic correlation.
}
\label{whole_curve}
\end{figure}

\begin{figure}
\includegraphics{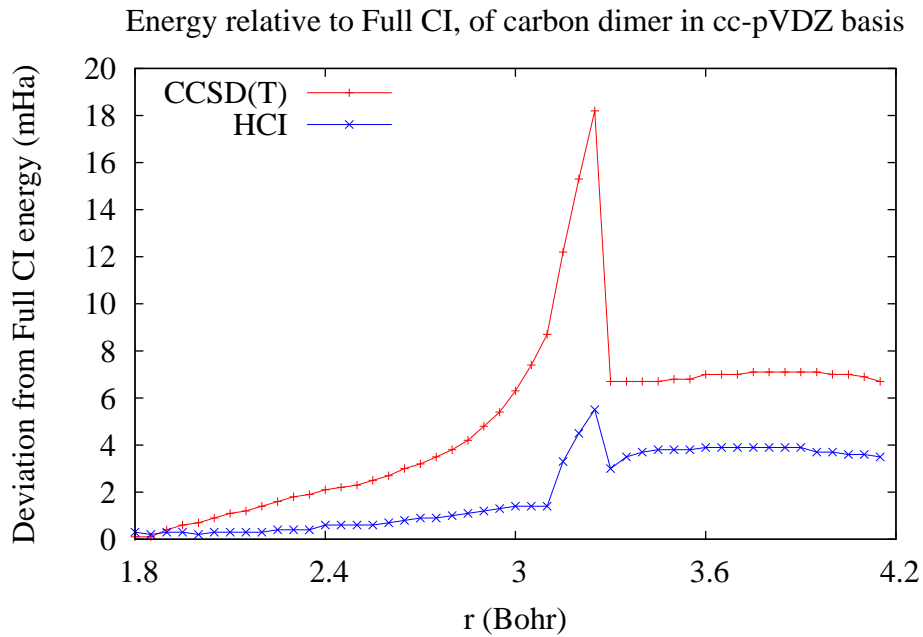}
\protect\caption{
Energy deviations relative to Full CI. HCI gets closer to the Full CI
energy, even in the region near equilibrium ($r=2.4$ Bohr). The discontinuity
near $r=3.25$ Bohr is due to a jump in the Hartree-Fock solution
due to a curve crossing between curves of different
symmetries~\cite{sharma2015multireference} and the fact that Hartree-Fock
is not guaranteed to find the global minimum~\cite{footnote}.
Both methods have difficulty in this region, but HCI has much less difficulty,
since its variational stage can describe the multireference character of the molecule.
}
\label{error}
\end{figure}

We compared to the ``gold standard" of quantum chemistry, coupled cluster
with singles, doubles, and perturbative triples (CCSD(T)).
As can be seen in Fig.~\ref{whole_curve}, HCI describes the whole binding curve
well, whereas CCSD(T) is only accurate near equilibrium. This shows that
HCI can capture both static and dynamic correlation effects. In Fig.~\ref{error},
we see that, even near equilibrium, where CCSD(T) is supposed to be good,
HCI still gets closer to the Full CI energy at these choices of $\epsilon_1$ and
$\epsilon_2$.

We used only $D_{2h}$ symmetry, rather than the
full $D_{\infty h}$ symmetry, so both coupled cluster and HCI
were less accurate in the region near $r=3.25$ Bohr, where a curve crossing
occurs between curves of different symmetries. This may explain why
HCI took $24\pm 4$ seconds to compute each energy in the range from $r=2.75$
to $3.30$ Bohr, while it only took $13\pm 1$ seconds to compute each energy outside
of that region.

\begin{figure}
\includegraphics{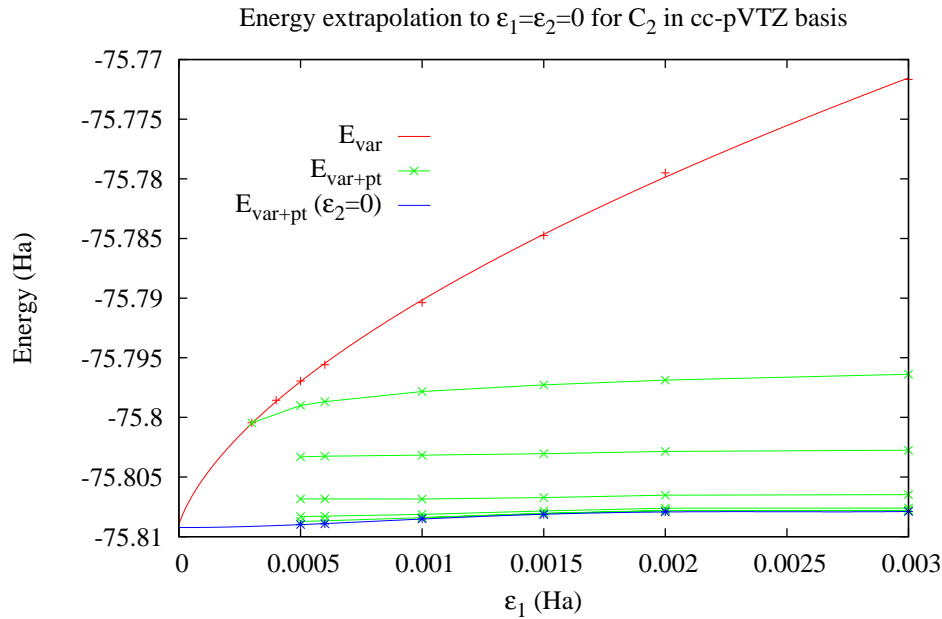}
\protect\caption{
Plot of the convergence to the Full CI limit ($\epsilon_1=\epsilon_2=0$) of the ground-state energy
of the carbon dimer at the equilibrium bond length of 1.24253 \AA\ in the cc-pVTZ basis set, with all electrons excited.
CASSCF(8,8) natural orbitals were used. The Full CI space for this system consists of about
$3\times 10^{14}$ Slater determinants.
The red points and line plot the variational energies, which depend only on
$\epsilon_1$. The green points and lines plot total energies (variational + perturbative correction)
for 5 different $\epsilon_2$ values ranging
from 3 to 300 $\mu$Ha (two of the lines
for the smallest values of $\epsilon_2$ are indistinguishable on the scale of the plot).
Finally, the blue points and line plot the total energies for different
values of $\epsilon_1$ at $\epsilon_2=0$
(i.e., the perturbative correction formula is extrapolated to obtain the exact limit).
The extrapolation function for the variational energies was chosen to be
a rational function of $\sqrt{\epsilon_1}$.
The total energies were extrapolated by first getting the total energy for each $\epsilon_1$ value
and $\epsilon_2=0$ using a rational polynomial in $\epsilon_2$ shown as blue stars,
and then fitting these values to a polynomial in $\epsilon_1$.
Our extrapolated ground state energy is -75.80924(15) Ha, in agreement with the value -75.809285
obtained from the DMRG calculation~\cite{olivaresamaya2015abinitio} with the largest bond dimension.
The lowest computed energy is -75.80873 for $\epsilon_1=5$ mHa and $\epsilon_2=3$ $\mu$Ha and the uncertainty in our extrapolated energy is given as $1/2$
of the energy extrapolation.
}
\label{vtz}
\end{figure}

In larger basis sets, the rate of convergence with respect to $\epsilon_1$ and $\epsilon_2$ can
be slower. For example, in Fig~\ref{vtz}, we plot the convergence of the ground state energy
of the carbon dimer in the cc-pVTZ basis at equilibrium, which has $3\times10^{14}$ determinants in the Full CI space.
Several of the energies for the smaller $\epsilon_1$ and $\epsilon_2$ values are within 1 mHa of the extrapolated energy
and these runs took between 4 and 9 minutes, compared to about 10 seconds for the cc-pVDZ basis.

\subsection{Chromium dimer}
We performed ground-state energy calculations on the challenging Cr$_2$ dimer in the Ahlrichs VDZ basis~\cite{schafer1992fully} at
$r=1.5$ \AA, both with frozen Mg core (24e, 30o), and with all electrons correlated
(48e, 42o). The Full CI spaces of these systems are approximately $9\times 10^{14}$ and
$2\times 10^{22}$ determinants, respectively.
We compared to extrapolated Density Matrix Renormalization Group (DMRG) results~\cite{olivaresamaya2015abinitio}.

We used natural orbitals from a (12e, 12o) CASSCF.
For both the 24- and 48-electron cases, we used $\epsilon_1=1$ mHa and
$\epsilon_2=10$ $\mu$Ha. As shown in table~\ref{cr2_table}, this choice in
parameters enabled HCI to produce energies within 1 mHa of the converged
DMRG results.

\begin{table}[htbp]
\begin{tabular}{cd{9}d{9}}
\hline
\hline
& \multicolumn{2}{c}{System}\\
\cline{2-3}
\multicolumn{1}{>{\centering\arraybackslash}m{23mm}}{Method}
    & \multicolumn{1}{>{\centering\arraybackslash}m{28mm}}{(24e, 30o)}
    & \multicolumn{1}{>{\centering\arraybackslash}m{28mm}}{ (48e, 42o)}\\
\hline
\hline
HCI  & -0.421 \, 30    & -0.444 \, 04     \tabularnewline
DMRG & -0.420 \, 95(3) & -0.444 \, 78(32) \tabularnewline
\hline
\end{tabular}
\caption{
Energies ($E$ + 2086 in Ha) of Cr$_2$ in the Ahlrichs VDZ basis at $r=1.5$ \AA, from
HCI($\epsilon_1=1$ mHa, $\epsilon_2=10$ $\mu$Ha) and converged DMRG~\cite{olivaresamaya2015abinitio}.
At these values of $\epsilon_1$ and $\epsilon_2$, the energies are within
1 mHa of the converged DMRG results.
}
\label{cr2_table}
\end{table}

At $\epsilon_1=1$ mHa, for (24e, 30o) and (48e, 42o) respectively, the variational wavefunctions consisted of $42\,945$ and $63\,592$ determinants,
with variational energies $-2086.368$ and $-2086.384$.

These HCI runs took only about two and eight
minutes for (24e, 30o) and (48e, 42o) respectively, on a single core.
This demonstrates that it is not significantly more challenging for HCI
to approximate the Full CI energy in this basis with all electrons correlated than with
core electrons frozen. We believe that this is because the Hamiltonian matrix elements
corresponding to core-valence excitations tend to be small (because of the small
overlap between sharp and diffuse orbitals), and HCI efficiently includes
only the most important excitations of that type.
However, note that whereas this Ahlrichs VDZ basis allows core-valence excitations,
it is not flexible enough to allow for core-core excitations,
so this conclusion may not generalize to basis sets such as Dunning's cc-pCVDZ
basis~\cite{dunning1989gaussian}.

\begin{figure}
\includegraphics{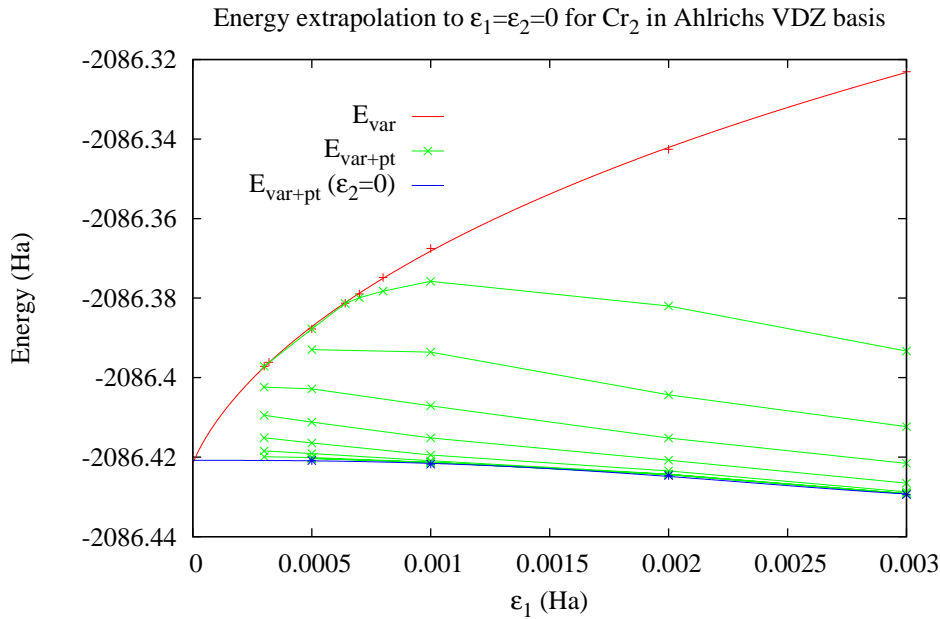}
\protect\caption{
Plot of the convergence of the ground-state energy of (24e, 30o) Cr$_2$ to the Full CI limit
($\epsilon_1=\epsilon_2=0$). 
The red and blue points and lines have the same meaning as in Figure~\ref{vtz}.
The green points and lines plot total energies (variational + perturbative correction)
for 9 different $\epsilon_2$ values ranging
from 2.5 to 640 $\mu$Ha, increasing by factors of 2 as one goes up the graph (several of the lines
for the smallest values of $\epsilon_2$ are indistinguishable).
Although the energies given in Table~\ref{cr2_table} were not extrapolated, this plot shows
how both the variational and total energies can be extrapolated to the 
Full CI limit.
The extrapolation functions were chosen to be of the same functional form as in Fig.~\ref{vtz}.
Our extrapolated ground state energy is -2086.4208(2) Ha, consistent with the extrapolated DMRG
energy given in Table~\ref{cr2_table}. The uncertainty in the extrapolated energy is given as $1/2$
of the energy extrapolation relative to the energy of -2086.42033 Ha obtained for $\epsilon_1=0.5$ mHa, $\epsilon_2=2.5$ $\mu$Ha.
This was most computationally-intensive run in this extrapolation plot and took about 14 minutes on a single core.
}
\label{extrap}
\end{figure}

A convergence plot of the ground-state energy with respect to the two input parameters $\epsilon_1$ and
$\epsilon_2$ is given in Fig.~\ref{extrap}. In the limit that $\epsilon_1=\epsilon_2=0$, the Full CI
energy is obtained, but highly-accurate results can be obtained much more cheaply at small but nonzero
values of $\epsilon_1$ and $\epsilon_2$.
Although the perturbative estimate is an underestimate of the true correction in Fig.~\ref{vtz}
and an overestimate in Fig.~\ref{extrap}, it is apparent from both figures that as the
variational wavefunction improves, the perturbative correction becomes progressively more effective
at recovering the missing energy.

For the converged runs that we did for C$_2$ in cc-pVDZ basis and Cr$_2$ in Ahlrichs VDZ basis,
the variational and perturbative stages both took approximately the same amount of time. However,
as previously mentioned, the perturbative stage was done in a na\"ive way that is space-limited.

\section{Outlook and ongoing research}
\label{ongoing_research}
We believe that the Heat-bath Configuration Interaction (HCI) algorithm described here
is an accurate and efficient method for treating many quantum many-body systems, including
several that are commonly considered to be strongly correlated.
There are many problems where the entire state space is enormous, but the portion of it that
makes a significant contribution to the exact many-body wavefunction is sufficiently small
that it can be included in the HCI wavefunction.
Since varying $\epsilon_1$ and $\epsilon_2$ enables different choices in the
speed/accuracy tradeoff, we believe that HCI will be competitive with many other electronic structure
methods, ranging from fast, approximate methods like Density Functional Theory (DFT), to highly-accurate,
expensive methods like Density Matrix Renormalization Group (DMRG) and Semistochastic Full Configuration Interaction
Quantum Monte Carlo (S-FCIQMC).
The most relevant competing method is DMRG.  For 3-dimensional systems, the bond dimension of DMRG scales
as $N^{2/3}$ so the computational cost scales exponentially in $N^{2/3}$ whereas the computational
cost of HCI is exponential in $N$.  Nevertheless, because HCI has a much smaller prefactor, we
expect that it will be the method of choice for some strongly correlated systems.

We are currently working on
extending HCI to be able to treat strongly-correlated systems in larger basis sets. One
promising idea in this vein is to
use subsets of the variational wavefunction at a time,
chosen either deterministically or stochastically
(using the stochastic efficient heat-bath sampling
algorithm~\cite{holmes2016efficient}). This not only
removes the storage bottleneck in our current implementation.

Computing the one- and two-body reduced density matrices of the variational
wavefunction is a trivial extension, and it will enable us to both perform orbital rotations
and compute ground-state molecular properties other than the energy.
Time-reversal symmetry and
$L_z$ symmetry for linear molecules
can also be used to reduce the effective Hilbert space size.
Extension to low-lying excited states is straight-forward using a method analogous
to that outlined in the original CIPSI paper~\cite{huron1973iterative}.
Finally, highly-accurate geometry optimizations can be performed, where early iterations
use fast, coarse estimates of the energy (large $\epsilon_1$, $\epsilon_2$), and later
iterations fine-tune those geometry configurations with more expensive runs (small $\epsilon_1$, $\epsilon_2$).

We are developing these extensions now.


%


{\it Acknowledgements:} 
We thank Garnet Chan and Sandeep Sharma for valuable discussions and Sandeep Sharma for 
help with using \textsc{Molpro} and understanding Cr$_2$ benchmarks and basis sets. 
This work was supported by grant NSF ACI-1534965.
NT was supported through the Scientific Discovery
through Advanced Computing (SciDAC) program funded by the U.S.
Department of Energy, Office of Science, Advanced Scientific Computing
Research, and Basic Energy Sciences.


\bibliographystyle{achemso}
\bibliography{hci}

\begin{mcitethebibliography}{23}
\bibitem[foo()]{footnote}
We used the D$_{2h}$ subgroup of the full D$_{\infty h}$ symmetry group for the
  two--electron integrals. Therefore the $^1\Sigma^+_g$ and $^1\Delta_g$ states
  of the D$_{\infty h}$ group can mix since they transform as the same
  irreducible representation of the subgroup. Hartree--Fock (HF) is not
  guaranteed to find the global minimum--energy determinant, and
  \textsc{Molpro} switches the occupied orbitals and the symmetry of the HF
  determinant from $^1\Sigma^+_g$ to $^1\Delta_g$ near $r=3.30$ Bohr if the HF
  calculations at each distance are done without providing input orbitals
  (although the $^1\Delta_g$ is the lower energy HF state for $r > 2.20$ Bohr).
  This results in a discontinuity in the Hartree--Fock energy, and smaller
  discontinuities in the CCSD(T) and HCI energies since these methods use the
  HF orbitals. Of course in the $\epsilon_1 \to 0$, $\epsilon_2 \to 0$ limit,
  the HCI energies would not have a discontinuity since the FCI energy is
  invariant to orbital rotations. Note that if the HF calculation was done
  using D$_{\infty h}$ symmetry then there would be only a derivative
  discontinuity at the curve crossing in the HF energy, but there would still
  be discontinuities in the CCSD(T) and HCI energies because of the
  discontinuity in the HF orbitals.\relax
\mciteBstWouldAddEndPunctfalse
\mciteSetBstMidEndSepPunct{\mcitedefaultmidpunct}
{}{\mcitedefaultseppunct}\relax
\EndOfBibitem
\end{mcitethebibliography}

\end{document}